\documentclass[sigplan,natbib=false,9pt,nonacm]{acmart}
\AtBeginDocument{%
  }

\copyrightyear{2026}
\acmYear{2026}
\setcopyright{none}
\acmConference[ICCAD '26]{IEEE/ACM International Conference on Computer-Aided Design}{November 08--12, 2026}{San Jose, CA, USA}
\acmBooktitle{IEEE/ACM International Conference on Computer-Aided Design (ICCAD '26), November 08--12, 2026, San Jose, CA, USA}
\acmDOI{10.1145/3831252.3834026}
\acmISBN{979-8-4007-2873-0/2026/11}

\usepackage{amsmath,amssymb,amsfonts}
\usepackage{algorithmic}
\usepackage{graphicx}
\usepackage{textcomp}
\usepackage{xcolor}
\usepackage{hyperref}
\usepackage{fontawesome5}
\usepackage{soul}
\usepackage{cuted}
\usepackage{capt-of} % Required to use captions outside of standard floats
\usepackage{dblfloatfix}

\usepackage[T1]{fontenc}
\usepackage[utf8]{inputenc}

\usepackage[
  backend=biber,
  style=ieee,
  citestyle=numeric-comp,
  sorting=none,
  giveninits=true,
  url=false,
  doi=false,
  isbn=false,
  eprint=false,
  maxnames=1,
  minnames=1
]{biblatex}

\AtEveryBibitem{%
  \clearlist{location}%
  \clearfield{address}%
  \clearlist{publisher}%
  \clearfield{month}%
  \clearfield{day}%
  \clearfield{note}%
  \clearfield{urldate}%
  \clearfield{series}%
}

\begin{document}
% \fontsize{9}{11}\selectfont

\title{TreeFI: Value-Aware Statistical Fault Injection for Deep Neural Networks
% \thanks{PEPR IA - Adapting project} MT: to protect double blind
}
\titlenote{\faGitlab\ \url{https://gitlab.inria.fr/nbires/treefi}}

\author{Noam Bires}
\email{noam.bires@inria.fr}
\affiliation[obeypunctuation=true]{%
  \institution{Univ Rennes, Inria, IRISA, CNRS},
  \city{Rennes},
  \country{France}
}
\author{Marcello Traiola}
\email{marcello.traiola@inria.fr}
\affiliation[obeypunctuation=true]{%
  \institution{Univ Rennes, Inria, IRISA, CNRS},
  \city{Rennes},
  \country{France}
}
\author{Angeliki Kritikakou}
\email{angeliki.kritikakou@inria.fr}
\affiliation[obeypunctuation=true]{%
  \institution{Univ Rennes, Inria, IRISA, CNRS},
  \city{Rennes},
  \country{France}
}
\author{Elisa Fromont}
\email{elisa.fromont@irisa.fr}
\affiliation[obeypunctuation=true]{%
  \institution{Univ Rennes, Inria, IRISA, CNRS},
  \city{Rennes},
  \country{France}
}

\begin{abstract}
Reliability evaluation of deep neural networks under hardware faults commonly relies on fault injection, but exhaustive campaigns are intractable for modern models and datasets. Statistical fault injection reduces this cost, yet existing approaches still require large injection budgets because they do not explicitly exploit a key property of floating-point faults: the effect of a bit flip depends strongly on the value being corrupted.

We propose \textbf{TreeFI}, a value-aware statistical fault-injection methodology for FP32 single-bit faults in DNN activations and weights. TreeFI partitions each layer's value distribution into intervals with similar expected bit-flip behavior, learned using regression trees, and allocates injections across these intervals according to their relevance for failure-rate estimation. This stratified allocation preserves the target confidence and error margin while avoiding unnecessary injections in low-impact regions of the fault space.

We validate TreeFI on CNN and Transformer models using CIFAR-10 and ImageNet. On ResNet8, where exhaustive activation fault injection is feasible, TreeFI provides more accurate estimates than state-of-the-art statistical FI baselines under the same campaign setting. Across the evaluated models, TreeFI reduces the required injection budget by up to 72.1$\times$, with average reductions of 44.9$\times$ for activation faults and 11.2$\times$ for the executed weight campaigns.
\end{abstract}


\maketitle

\section{Introduction}
Deep Neural Networks (DNNs) are increasingly deployed in safety- and mission-critical systems, where dependability is as important as predictive performance. In these domains, transient hardware faults, such as soft errors induced by radiation or electrical noise, may cause bit flips that silently corrupt intermediate computations and alter the final prediction, affecting the system's correctness. Assessing the impact of such faults is therefore necessary to evaluate the reliability of the system~\cite{bolchiniResilienceDeepLearning2024,ahmadilivaniSystematicLiteratureReview2024,ruospoSurveyDeepLearning2023a,rech2024reliability,su2023testability}.

Because the effects of these faults depend on where and when they occur, reliability evaluation typically relies on fault injection (FI), which emulates faults during execution and observes their impact on the model output~\cite{ruospoSurveyDeepLearning2023a,rech2024reliability}. However, exhaustive FI rapidly becomes impractical for modern model architectures and datasets, because it requires exploring a massive space of injection locations within the model, bit positions, and inputs. To address this challenge, Statistical Fault Injection (SFI) approaches have been proposed to reduce the cost of FI, while still providing statistically meaningful reliability estimates~\cite{leveugleStatisticalFaultInjection2009,ruospoAssessingConvolutionalNeural2023a,ruospoEffectiveIterativeStatistical2025a}. These approaches significantly improve scalability compared with exhaustive campaigns and currently represent the state-of-the-art for efficient FI evaluation of neural networks~\cite{ruospoAssessingConvolutionalNeural2023a,ruospoEffectiveIterativeStatistical2025a}.
Despite these advances, current SFI approaches still neglect an important source of variation in the fault space: the effect of a bit flip strongly depends on the value being corrupted. This is particularly important when the values in a DNN layer span a wide range. For a specific bit position, the original value determines both whether this bit is 0 or 1 and how much the value changes when this bit is flipped.
As a result, two injections targeting the same bit position can have very different effects depending on the stored value.
However, existing SFI methods do not systematically exploit this behavior, leaving room for further reductions in the number of required injections.

To address this gap, this paper proposes \texttt{TreeFI}, a value-aware statistical fault injection methodology for evaluating DNN reliability. The key idea is simple: bit flips do not have the same impact for all values, so TreeFI first groups layer values into ranges that behave similarly under bit flips. It then allocates more injections to the value ranges that matter most for estimating the failure rate, while still targeting a user-specified confidence and error margin. To do so, TreeFI learns value ranges for each layer and bit position using regression trees, and then derives a stratified sampling strategy that requires substantially fewer injections than state-of-the-art SFI approaches.
This work adopts the single-bit fault model commonly used in statistical fault injection studies of deep neural networks (DNNs)~\cite{ruospoAssessingConvolutionalNeural2023a,ruospoEffectiveIterativeStatistical2025a}, where faults are represented as independent bit corruptions under the single-fault assumption.
The objective of this work is not to evaluate the realism of the fault model, but to \textbf{reduce the number of injections required to achieve a target statistical accuracy} given a predefined fault population. Software‑level single‑bit injections are therefore used solely to define a fully characterizable reference population, enabling rigorous validation and comparison of the proposed statistical acceleration method.
Although real hardware faults may produce multi-bit errors or exhibit temporal and spatial correlations, accurately modeling such effects requires detailed low-level hardware characterization and is beyond the scope of this study.

We evaluate TreeFI using software FI, in line with prior SFI literature, and compare it against state-of-the-art data-aware one-shot~\cite{ruospoAssessingConvolutionalNeural2023a} and iterative Statistical FI~\cite{ruospoEffectiveIterativeStatistical2025a} approaches. We first evaluate the proposed methodology on two Convolutional Neural Network (CNN) architectures: ResNet8~\cite{heDeepResidualLearning2016} and RepVGG-A0~\cite{dingRepVGGMakingVGGstyle2021} on the CIFAR-10~\cite{krizhevskyLearningMultipleLayers} dataset. To establish a ground-truth reference, we perform an exhaustive FI campaign for ResNet8 activations on 5{,}000 inputs and use it to evaluate the quality of the resulting reliability estimates. We then study the scalability of the proposed methodology using a much larger dataset, i.e.,  ImageNet~\cite{dengImageNetLargescaleHierarchical2009}, with different Transformer architectures, such as DeiT-Tiny/Small/Base~\cite{touvronTrainingDataefficientImage2021}.
Overall, the results show that TreeFI achieves up to a \textbf{72.1$\times$} reduction in injections compared to the data-aware one-shot SFI. Moreover, on the ResNet8 benchmark, where exhaustive ground-truth FI is possible, TreeFI yields lower estimation error in a single statistical FI campaign than the data-aware one-shot SFI.
In summary, the main contributions of this work are:

\noindent $\bullet$~a value-aware SFI methodology that exploits the
    dependence of bit-flip effects on the corrupted value;
    %value-dependence of transient bit flips;

\noindent $\bullet$~a per-layer, per-bit partitioning of value ranges into intervals with similar bit-flip behavior, learned using a regression tree;

\noindent $\bullet$~a stratified sampling strategy that targets a desired confidence level and error margin, and reduces the required injections;
    % \item experimental evaluations comparing TreeFI with exhaustive FI, possible only for small scale networks such as XXX, and two state-of-the-art SFI approaches, including a hyperparameter ablation study on ResNet8, an activation study on DeiT models using ImageNet, and a weight study on CNN models with additional preprocessing-derived count estimates for DeiT.

\noindent $\bullet$~a thorough experimental evaluation that validates TreeFI against exhaustive FI (ResNet8), compares it with two state-of-the-art statistical FI baselines, and includes a hyperparameter ablation study.
% \end{itemize}

\section{Background and related work}
\label{sec:related}
Deep neural network reliability has been studied across multiple abstraction levels, from software-implemented fault injection to hardware-level simulation and radiation-based experiments. Recent surveys organize this literature according to the goal of the work (reliability assessment versus hardening), the abstraction level of the analysis, the target platform, and the adopted fault model \cite{bolchiniResilienceDeepLearning2024,ahmadilivaniSystematicLiteratureReview2024,ruospoSurveyDeepLearning2023a,rech2024reliability,su2023testability}. Radiation-based fault injection experiments provide high realism, but they require access to dedicated irradiation facilities and the target device, which makes them costly and less flexible to repeat. RTL- or architecture-level fault injection can capture hardware details more faithfully, but it depends on the availability of low-level hardware models and is typically much more expensive to run.
By contrast, software-implemented fault injection, while less accurate, is hardware-agnostic and can be applied early in the development cycle, before the final accelerator or detailed RTL model is available. This makes it a practical setting for studying faults in weights, activations, and intermediate computations while keeping the experimental campaign manageable. A handful of software frameworks support this style of analysis, including Ares~\cite{reagenAresFrameworkQuantifying2018b}, TensorFI~\cite{chenTensorFIFlexibleFault2020}, PyTorchFI~\cite{mahmoudPyTorchFIRuntimePerturbation2020a}, and MRFI~\cite{huangMRFIOpenSourceMultiresolution2024}.

To reduce the computational cost of fault-injection campaigns, \textit{statistical fault injection (SFI)} techniques are widely used. Instead of exhaustively injecting all possible faults, SFI injects only a carefully selected subset and uses classical sampling theory to estimate the behavior of the full fault population under a desired confidence level and error margin~\cite{cochran1977sampling,leveugleStatisticalFaultInjection2009}. In the DNN setting, these methods aim to make reliability analysis scalable while remaining consistent with exhaustive FI under the same fault model~\cite{ruospoAssessingConvolutionalNeural2023a,ruospoEffectiveIterativeStatistical2025a}. In practice, they organize the statistical process at the layer and bit level, and determine how many injections are needed for each layer--bit pair.

State-of-the-art methods in this line include the data-aware one-shot method~\cite{ruospoAssessingConvolutionalNeural2023a} (\texttt{SFI}) and the later iterative method~\cite{ruospoEffectiveIterativeStatistical2025a} (\texttt{IFI}). \texttt{SFI} reduces the number of injections through proxy-guided allocation, but it was originally proposed for weight fault injection and is less favorable on activation faults, where the value distribution is more heterogeneous and input-dependent. \texttt{IFI} updates the campaign using intermediate FI results, but this comes with two practical limitations: it can spend many injections in early iterations on bits that ultimately have little or no impact, and its iterative nature makes the total runtime difficult to predict in advance.

TreeFI follows this statistical-FI line, but introduces an additional source of structure that these methods do not explicitly exploit: the dependence of fault effects on the corrupted value. It partitions the value space into intervals with similar bit-flip behavior and allocates more injections to the intervals that matter most for the failure rate estimation. In this way, TreeFI further reduces the number of required injections, while keeping the advantages of one-shot, upfront-budgeted campaigns. Our approach is validated using software fault injection, in line with state-of-the-art methods.

%AK: we should improve next paragraph to support more the need of SFI
Other approaches reduce the cost of FI through techniques that are complementary to statistical fault injection. Some works target \emph{critical-bit or parameter identification}, for example through binary-search-like fault localization~\cite{chenBinFIEfficientFault2019}, bit-level sensitivity analysis~\cite{wengFKerasSensitivityAnalysis2024}, Hessian-guided ranking combined with adaptive FI~\cite{wengPrioriFIMoreInformed2026}, or machine-learning-guided ranking of vulnerable components~\cite{traiolaMachinelearningguidedFrameworkFaulttolerant2023,ETS-2023}. These methods aim to identify the most sensitive parts of the model and focus the campaign on them, rather than estimating the full failure rate under an explicit confidence/error target. Other approaches replace or accelerate FI through analytical or hybrid modeling, such as semi-analytical vulnerability estimation~\cite{ahmadilivaniDeepVigorScalableAccurate2025} or cross-layer simulation~\cite{Bolchini-xlayer,tonettoENFORSAEndtoendCrosslayer2026}, where only the most hardware-critical parts are simulated at low level while the rest of the system is evaluated at a higher level of abstraction. TreeFI is complementary to these directions: rather than replacing FI with analytical approximations or prioritizing only the most critical components, it improves fault-sampling efficiency in statistical FI by explicitly leveraging value dependence.

Other studies have shown that resilience depends strongly on numerical format and value range, but they use this observation for protection rather than vulnerability assessment. These works include format-aware resilience analysis~\cite{gutierrez-zaballaEvaluatingSingleEvent2024}, value-range constraining techniques such as numerical-range squeezing~\cite{ozenSNRSqueezingNumerical2021}, and reliability-oriented training methods based on trainable activation shaping or transient-fault-aware optimization~\cite{ghavamiFitActErrorResilient2022a,santosImprovingDeepNeural2025}. By contrast, our focus is not to harden the model, but to evaluate its vulnerability more efficiently under a statistical FI methodology.

\begin{figure*}[!t]
    \centering
    \includegraphics[width=.8\textwidth,trim={10 10 10 10},clip]{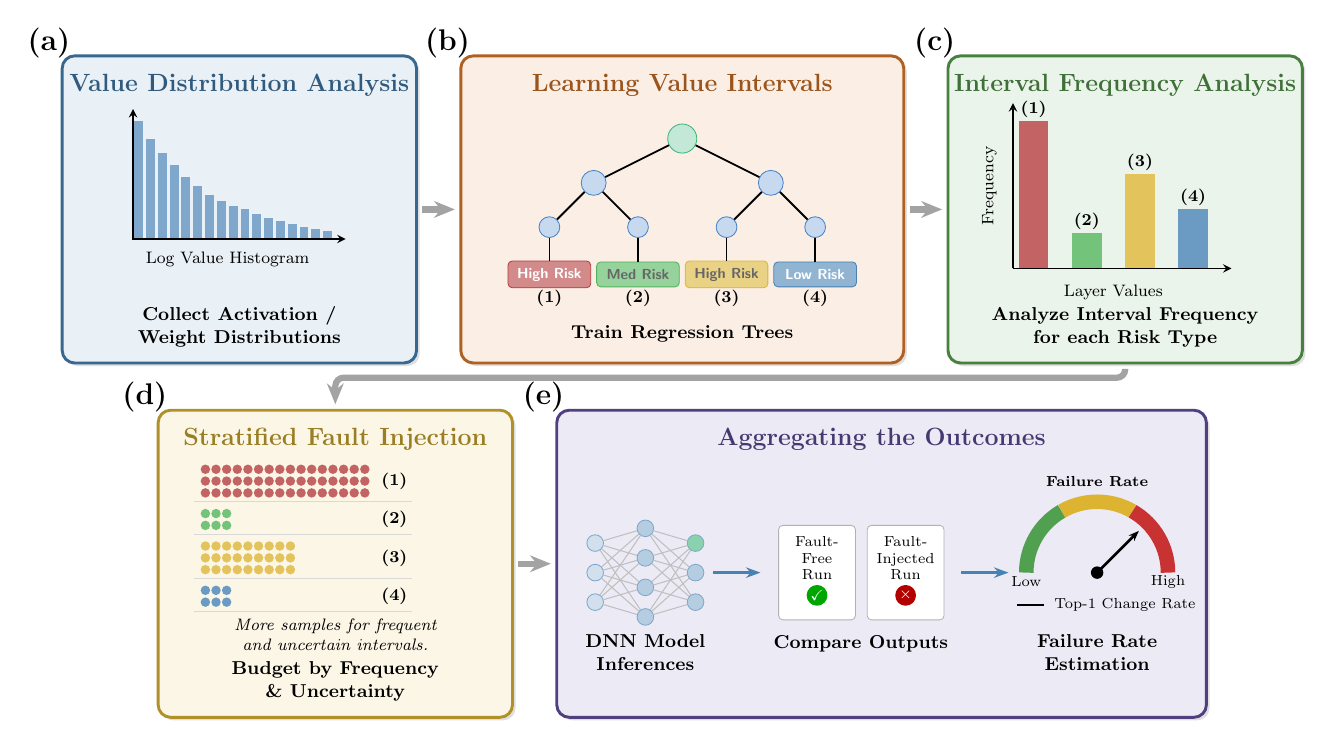}
    \captionof{figure}{Overview of the TreeFI Value-Aware Fault Injection Method.}
    \label{fig:fault_injection}
\end{figure*}

\begin{figure}[b]
\vspace{-5pt}
    \centering
    \includegraphics[width=.7\columnwidth]{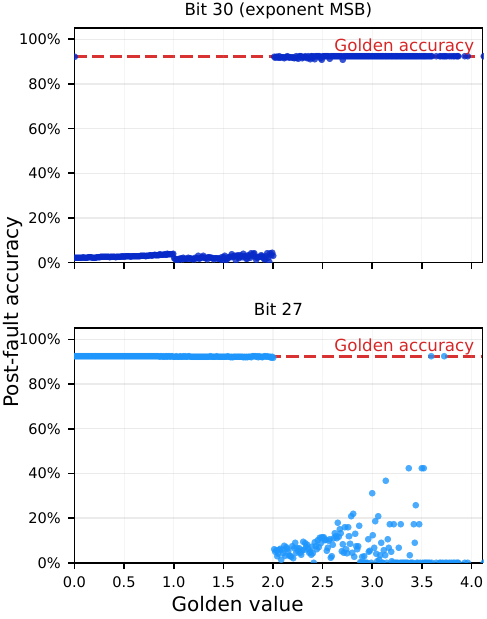}
    \vspace{-5pt}
    \captionof{figure}{Motivating example on a representative ResNet8 activation layer: for a fixed bit position, the impact of a bit flip on model accuracy depends strongly on the original activation value.}
    \label{fig:motivation}
\end{figure}
% \addtocounter{figure}{1}

\section{TreeFI: Value-aware statistical fault injection}

\subsection{Motivation}
A key limitation of existing statistical fault injection methods is that they do not explicitly organize the fault sampling campaign by value range. In practice, however, the impact of a bit flip depends not only on the bit position but also on the original numerical value. This dependence is especially pronounced for floating-point data, where flipping the same bit can produce very different perturbations depending on the value being corrupted.
\autoref{fig:motivation} illustrates this effect on a representative ResNet8 activation layer. Each plot is based on 65{,}536 single-bit fault injections. In both plots, the x-axis represents the original activation value, and the y-axis represents the model accuracy after flipping bit 30 (top graph) or bit 27 (bottom graph) of that activation. The two subplots correspond to two different bit positions, but the main message is the same in both cases: for a given bit position, the effect of the bit flip is not uniform, but depends on the original value before the flip. As shown in the figure, for some activation ranges, flipping the bit has little or no visible effect on model accuracy, whereas in other ranges the same bit causes a strong drop in accuracy. Therefore, two fault injections targeting the same layer and the same bit position can still produce very different outcomes because the corrupted values belong to different value ranges. The same value-dependent behavior is also observed for weight faults, although \autoref{fig:motivation} focuses on activations.

This observation motivates our approach. If the fault effect depends strongly on the corrupted value, then a sampling strategy that does not explicitly distinguish between value ranges is inefficient. TreeFI's central idea is to exploit the value dependence highlighted in \autoref{fig:motivation}. While existing statistical FI approaches already allocate injections across layers or bit positions, they do not explicitly organize the sampling campaign by value range. TreeFI addresses this limitation by partitioning the value distribution into intervals and allocating more injections to those that occur more frequently and are more likely to influence the final failure-rate estimate. Thus, TreeFI uses the value axis itself to guide the sampling campaign, making it more efficient while still targeting a desired confidence level and error margin.

\vspace{-5pt}
\subsection{Overview of TreeFI}

The proposed value-aware fault-injection methodology comprises two phases, as depicted in \autoref{fig:fault_injection}. The first phase is an \emph{offline characterization}, corresponding to \autoref{fig:fault_injection}(a)--(c). TreeFI first collects the value distribution of each target layer (a). It then partitions the value axis into intervals with similar local bit-flip effects (b), based only on the numerical difference between the original value and the bit-flipped value, without evaluating the effect on the model output. Finally, it computes how frequently each interval occurs in practice (c). The second phase performs the actual fault injection campaign, as shown in \autoref{fig:fault_injection}(d)--(e). In this phase, TreeFI determines how many injections to apply in each interval and performs the corresponding fault injections (d), and finally aggregates the observed results to obtain a final failure-rate estimate (e).
The same workflow applies to faults in both activations and weights, but the practical execution differs slightly between the two cases. The effect of a fault at a given activation tensor location depends on its value, which varies across input samples. As a result, selecting an activation fault requires considering both the input and the activation location. For weight faults, the value to be corrupted is stored in the model parameters and is independent of the input. Once a weight and bit position are selected, the same corrupted parameter can be evaluated across all considered inputs.
The next two subsections describe the different phases of the TreeFI approach.

\subsection{TreeFI phases}
\label{sec:offline_char}

\textbf{\textit{(a) Value Distribution Analysis}}
The first step, sketched in \autoref{fig:fault_injection}(a), is to collect the value distribution for each target DNN layer.
For activations, TreeFI executes the model with the validation set and records the values at the output of the layer. For each input, all tensor values of the target layer are accumulated into the histogram; no tensor-value subsampling is used. To keep preprocessing tractable, TreeFI stores only weighted histogram counts and discards the raw activation values after they have been accumulated.
For faults in weights, the distribution is obtained directly from the model parameters.
TreeFI achieves lightweight preprocessing by storing these values in histogram form rather than in a large list of raw values. Indeed, in large-scale settings, direct raw-value sorting can require more than one day of preprocessing, whereas our histogram-based approach reduces this cost to a few seconds. To obtain the histogram, values that differ by less than $10^{-4}$ are grouped together in bins, and the frequency of each bin is recorded. In our experiments, using this approach produces identical interval partitions (phase b) to those obtained from the raw values.
All subsequent partitioning steps are weighted by bin counts, ensuring that rare values do not disproportionately influence the resulting intervals.
The histogram-based approach and the $10^{-4}$ threshold are used only to speed up offline characterization; they do not define the precision of the injected values. The \textit{original value} of the activation or weight, $a$, is used for subsequent fault injection and failure-rate estimation.

\textbf{\textit{(b) Learning Value Intervals}}
The second step, shown in \autoref{fig:fault_injection}(b), is to partition the values collected in the first step into intervals with similar local bit-flip effects. Here, ``local'' means that TreeFI does not yet perform full fault injections through the DNN or evaluate the effect on the model output. Instead, for each layer and bit position, it only considers how much the value itself changes when that bit is flipped.

For each layer and bit position, TreeFI learns the intervals with a one-dimensional regression tree trained on the collected value histograms. Intuitively, the regression tree scans the value axis and places split points where the local effect of flipping the selected bit changes significantly. The histogram counts from step~(a) are used during training so that the learned splits are supported by enough observed values, rather than being driven by rare bins with very few samples. As a result, each leaf of the tree defines one interval.
In our experiments, trees are trained independently for each layer--bit pair using a minimum leaf weight of 10 samples, a maximum of 100 leaf nodes, and a weighted-MSE split threshold of $10^{-4}$. The maximum-leaf constraint is not reached in practice: the learned trees contain at most about 15 leaves per layer--bit pair.

For each learned interval, TreeFI then computes an average severity based on the log-distance between the original value and its bit-flipped counterpart, averaged over the values in that interval. We denote this quantity by $\textit{avg\_log}_{\ell,b,k}$. This severity is mapped through a sigmoid function $\sigma$ to obtain the \emph{risk score} $r_{\ell,b,k}$ of interval $k$ in layer $\ell$ for bit position $b$:
\begin{equation}
r_{\ell,b,k} = \sigma\bigl(s\cdot(\textit{avg\_log}_{\ell,b,k} - \beta_\ell)\bigr),
\end{equation}
where $s$ controls how sharp the transition is and $\beta_\ell$ is a layer-dependent bias. In our implementation, $s$ is fixed across all layers, while the bias is defined as
\begin{equation}
\beta_\ell = \log_{10}(\gamma\cdot a_{\ell,\max}),
\end{equation}
where $a_{\ell,\max}$ is the maximum golden (i.e., original non-faulty) value observed in layer $\ell$ and $\gamma$ is a scaling factor. The resulting \emph{risk score} is bounded between 0 and 1 and is used later to guide the distribution of fault injections across intervals.

These intervals serve as the basic units for fault sampling and estimation. This step converts the qualitative observation of \autoref{fig:motivation} into an operational fault-injection strategy.

\textbf{\textit{(c) Interval Frequency Analysis}}
Once the value intervals are defined, TreeFI computes how often each interval appears during the golden execution, as illustrated in \autoref{fig:fault_injection}(c). This interval frequency comes directly from the histogram collected in step~(a) and indicates how common each interval is during fault-free execution. These frequencies are used in two ways: first, decide how many fault injections to assign to each interval (d), and second, combine the interval-level results into a final failure-rate estimate (e).

\textbf{\textit{(d) Stratified Fault Injection}}
The interval risk scores computed in step~(b) are used here to capture how much the expected fault effect varies within each interval. This step then decides how many fault injections to assign to each value interval, as sketched in \autoref{fig:fault_injection}(d). For each layer and bit position, TreeFI distributes injections across the learned intervals using two pieces of information: how often each interval appears during fault-free execution, and how much the expected bit-flip effect varies within that interval. Following the standard idea of stratified sampling~\cite{neymanTwoDifferentAspects1992}, intervals that are more common and intervals whose effect is harder to estimate receive more injections. In our case, the number of injections  $n_{\ell,b,k}$ assigned to interval $k$ is proportional to
\begin{equation}
n_{\ell,b,k} \propto W_{\ell,b,k}\sqrt{r_{\ell,b,k}\bigl(1-r_{\ell,b,k}\bigr)},
\end{equation}
where $W_{\ell,b,k}$ is the frequency of interval $k$ for a given pair of layer ($\ell$) and bit ($b$) and $r_{\ell,b,k}$ is the interval risk score. The values $n_{\ell,b,k}$ directly give the number of injections assigned to each interval. Their sum over $k$ gives the total number of injections for the considered layer--bit pair, and this total is chosen so that the final failure-rate estimate satisfies the desired confidence level and error margin.

As a result, TreeFI does not distribute injections uniformly across the value axis. Instead, it concentrates more injections on the value intervals that are both common and more difficult to estimate accurately. In this way, TreeFI addresses the limitation highlighted in \autoref{fig:motivation}: since the effect of a bit flip varies across the value axis, the number of injected faults should also vary across value intervals.

\textit{Activation and weight execution:}
For activation faults, TreeFI processes each input and looks for activation values whose fault-free value falls inside the selected interval. Among these candidate locations, it randomly selects one location and flips the chosen bit. In other words, an activation location is eligible for injection if its value, for the current input, belongs to the interval selected for sampling.
For weight faults, TreeFI selects a model parameter whose value falls inside the selected interval and flips the chosen bit in that parameter. The model is then executed with this corrupted parameter on the considered input samples.
In both cases, the injected event in the current implementation is a single-bit flip. We use this fault model because it is standard in software-implemented reliability studies and provides a simple and controlled setting for evaluating the proposed sampling methodology and comparing it to existing approaches, which use the same fault model~\cite{ruospoAssessingConvolutionalNeural2023a, ruospoEffectiveIterativeStatistical2025a}.

\textbf{\textit{(e) Aggregating the outcomes}}
The final step is shown in \autoref{fig:fault_injection}(e). For each fault injection run, TreeFI compares the output of the faulty execution with the output of the fault-free (golden) execution and records whether the top-1 prediction changes. Repeating this process for all injections assigned to a given interval produces a measured failure rate for that interval, i.e., the fraction of injections in that interval that change the prediction.

To estimate the failure rate for one layer--bit pair, TreeFI combines the measured failure rates of all intervals using how often these intervals appear during the golden pass:
\begin{equation}
\hat{p}_{\ell,b} = \sum_k W_{\ell,b,k}\,\hat{p}_{\ell,b,k},
\end{equation}
where $\hat{p}_{\ell,b}$ is the estimated failure rate for layer $\ell$ and bit position $b$, $W_{\ell,b,k}$ is the frequency of interval $k$ for that layer--bit pair, and $\hat{p}_{\ell,b,k}$ is the measured failure rate obtained from the injections performed in interval $k$.
This weighted combination is a key part of TreeFI. It ensures that the final failure-rate estimate reflects the value distribution observed during fault-free execution. Without this step, all intervals would contribute equally, even though some occur much more often than others. By weighting each interval with its empirical frequency, TreeFI makes the final estimate more representative of the actual layer behavior. This is one of the main differences between TreeFI and statistical FI methods that do not combine results at the value-interval level.

\section{Experimental set-up and evaluation}
\label{sec:results}
% \subsection{Experimental set-up}
We evaluate TreeFI on image classification DNN models under bit flips in activations and weights. We compare the proposed approach against exhaustive fault injection when this is computationally feasible, and against two state-of-the-art statistical FI baselines: the data-aware one-shot method SFI~\cite{ruospoAssessingConvolutionalNeural2023a} and the iterative method IFI~\cite{ruospoEffectiveIterativeStatistical2025a}.
For CIFAR-10 experiments, ResNet8 and RepVGG-A0 are trained on the standard 50{,}000-image training split and fault-injection campaigns are evaluated on images from the separate 10{,}000-image test split, which is not used during training. For the DeiT experiments, we use ImageNet-pretrained checkpoints and evaluate fault injection on images from the ImageNet validation set, which is distinct from the training set. Thus, the inputs used for validation and fault injection are unseen during model training.
In our experimental set-up, statistical campaigns target an absolute error margin of $E=1\%$ with $99\%$ confidence. IFI follows the recommended protocol of initializing with a looser target ($E=5\%$) and refining to $E=1\%$. For TreeFI, we report results with $\gamma \in \{1,2,5,10\}$ for the layer-dependent bias defined in Section~\ref{sec:offline_char}, while keeping the sigmoid sharpness fixed at $s=2$ across all layers.
All fault injection campaigns reported in this paper were executed on an NVIDIA H100 GPU with 96\,GB of VRAM.

This work instantiates the methodology using FP32 single-bit flips during inference. However, the proposed approach is not tied to FP32 specifically. More generally, it applies whenever the fault model defines how a bit flip modifies a value and the corresponding value distribution can be collected and partitioned into intervals.
To evaluate the statistical robustness of TreeFI, we run 15 independent TreeFI campaigns on ResNet8 activations with different random seeds. Using the exhaustive ResNet8 reference, we report both the estimation error and the variability across runs. This allows us to evaluate how close the estimated failure rates are to the exhaustive reference and how stable the method is across repeated campaigns.

We begin with activation faults on CIFAR-10 using ResNet8, where we also run an exhaustive fault injection campaign. This exhaustive campaign provides a reference against which the reliability estimates of TreeFI, SFI, and IFI can be directly compared. We then extend the activation-fault evaluation to RepVGG-A0 on CIFAR-10 and to larger transformer models from the DeiT family on ImageNet, in order to study the scalability of the proposed approach. Since exhaustive FI is infeasible at the ImageNet scale, the transformer analysis focuses on statistical fault injection results only. We also present an ablation study on ResNet8 activations to assess the robustness of the proposed scoring parameters. We then report results from fault-injection campaigns targeting DNN weights.
For weights, we executed all-layer campaigns on ResNet8, 10-layer campaigns on RepVGG-A0, and 6-layer campaigns on DeiT-Tiny with IFI, SFI, and TreeFI-$20\times$.

\subsection{Faults in activations}

\subsubsection{Validation against exhaustive fault injection (ResNet8 - CIFAR-10)}
\begin{figure}[!t]
  \centering
  \includegraphics[height=0.5\textwidth]{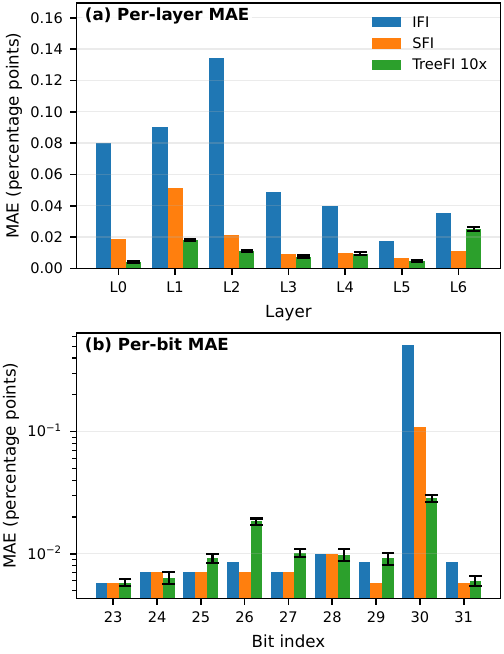}
  \vspace{-7pt}
  \caption{ResNet8 MAE to exhaustive activation failure rates: (a) per-layer MAE over bits 23--31 and (b) per-bit MAE over layers, for IFI, SFI, and TreeFI-$10\times$ (b uses log scale).}
  \label{fig:resnet8_mae}
  \vspace{-8pt}
\end{figure}

We compare TreeFI and SFI against exhaustive FI for faults in ResNet8 activations, since this benchmark is still small enough to allow a full exhaustive campaign.
TreeFI-$10\times$ (i.e., $\gamma = 10$) obtains failure rates that are very similar to those of the exhaustive reference, reproducing the same most vulnerable bits and the layers with the highest failure rates.
This indicates that the reduction in the number of injections, shown in the remainder of the text, does not come from overlooking sensitive cases, but from distributing injections more effectively across the value space.
\autoref{fig:resnet8_mae} reports the mean absolute error (MAE) of TreeFI-$10\times$, SFI, and IFI, with respect to the exhaustive failure rates. Especially around the most failure-prone layer and bit regions (i.e., L0-L2, and bit 30), TreeFI achieves lower MAE than both SFI and IFI while requiring substantially fewer injections. In general, the reported MAE values are small, and all are within the 1\% statistical error target.  In particular, TreeFI achieves lower per-layer MAE because it computes the interval-level failure rates using the true interval frequencies observed during the golden pass, so that each interval contributes to the final estimate according to how often it actually occurs. By contrast, SFI and IFI do not explicitly reweight the estimate in this way. As a result, TreeFI better captures the contribution of value ranges that occur less often but can still strongly affect the final failure rate. In other words, the reduction in the number of injections does not come at the expense of estimate quality.

When varying $\gamma$, the MAE variations among TreeFI configurations are also small in absolute terms: the gap between TreeFI-$1\times$ and TreeFI-$10\times$ stays below $0.05$ percentage points across all layers and bits, which is one order of magnitude smaller than the target error margin of $1\%$.
The measured runtimes follow the same trend as the reduction in the number of injections reported in \autoref{tab:inj_counts_act}. On ResNet8 activations, exhaustive FI requires about $270\times$ more injections than TreeFI-$10\times$, highlighting the scale of the reduction achieved by the proposed approach. Among the statistical methods, SFI requires 740{,}354 injections and 7\,h\,13\,min, IFI requires 979{,}021 injections and 9\,h\,35\,min, and TreeFI-$10\times$ requires only 34{,}934 injections and 28\,min. This corresponds to a theoretical reduction of about $21.2\times$ versus SFI and $28.0\times$ versus IFI, while the measured wall-clock reduction is about $15.5\times$ versus SFI and $20.5\times$ versus IFI. Therefore, the practical runtime savings follow the same overall trend as the reduction in the number of injections.

\subsubsection{Efficiency}
\begin{figure}[!b]
\centering
\vspace{-5pt}
\includegraphics[width=\columnwidth]{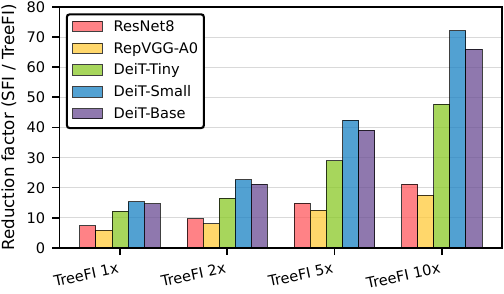}
\vspace{-9pt}
\caption{Activation-injection reduction factors relative to the SFI baseline. Values $>1$ indicate fewer injections than SFI.}
\label{fig:inj_reduction_vs_sfi}
\end{figure}

Overall, we observe that TreeFI reduces the number of required fault injections, while keeping the same confidence and error margin, across all models.
\autoref{tab:inj_counts_act} reports the absolute number of activation fault injections required by each method under the same statistical objective. For DeiTs and IFI, we do not report the number of injections since, as already pointed out in Section~\ref{sec:related}, IFI's iterative nature makes the total number of faults impossible to predict in advance, and running a full-model analysis with IFI for DeiTs is not tractable in our setup.
\autoref{fig:inj_reduction_vs_sfi} presents the reduction factor achieved by TreeFI relative to SFI (values $>1$ indicate fewer injections).
Across all models, TreeFI consistently reduces the number of required injections, ranging from $5.9\times$ fewer injections (RepVGG-A0, $\gamma=1\times$) up to $72.1\times$ fewer injections (DeiT-Small, $\gamma=10\times$). On average, TreeFI-$10\times$ achieves the same statistical results as SFI (see next paragraph), with 44.9$\times$ fewer injections.
Since the total FI runtime scales approximately linearly with the number of injections, the achieved reductions translate directly into faster fault injection campaigns.

\begin{table}[tb!]
\centering
\caption{Total number of required injections on \textbf{activations} per method and per model.}
\vspace{-5pt}
\label{tab:inj_counts_act}
\setlength{\tabcolsep}{2pt} % Default value: 6pt
\resizebox{\columnwidth}{!}{
\begin{tabular}{lrrrrr}
\hline
Method & \textbf{ResNet8} & \textbf{RepVGG-A0} & \textbf{DeiT-Tiny} & \textbf{DeiT-Small} & \textbf{DeiT-Base} \\
\hline
Exhaustive & 9{,}437{,}184 & 14{,}024{,}704 & 58{,}097{,}664 & 116{,}195{,}328 & 232{,}390{,}656 \\
SFI \cite{ruospoAssessingConvolutionalNeural2023a} & 740{,}354 & 1{,}812{,}919 & 1{,}664{,}695 & 1{,}749{,}489 & 1{,}781{,}928 \\
IFI \cite{ruospoEffectiveIterativeStatistical2025a} & 979{,}021 & 2{,}173{,}770 & N/A & N/A & N/A \\
\hline
TreeFI 1$\times$  & 100{,}046 & 308{,}303 & 137{,}091 & 113{,}820 & 121{,}702 \\
TreeFI 2$\times$  & 76{,}810  & 227{,}601 & 100{,}891 & 77{,}458 & 84{,}379 \\
TreeFI 5$\times$  & 49{,}840  & 145{,}525 & 57{,}519  & 41{,}291 & 45{,}672 \\
TreeFI 10$\times$ & 34{,}934  & 103{,}552 & 34{,}816  & 24{,}254 & 26{,}970 \\
\hline
\end{tabular}
}
\vspace{-5pt}
\end{table}

Note that the gains of TreeFI become more pronounced as model size increases, as illustrated by the DeiT family. This trend suggests that the benefits of the proposed value-aware approach grow with the size of the fault space and with the heterogeneity of the underlying value distributions and fault effects. As the number of possible fault injections increases, existing methods, that do not explicitly organize sampling by value range, spend more injections on less impactful values, whereas TreeFI concentrates the campaign on intervals that are more likely to influence the final reliability estimate.
As a result, TreeFI is particularly attractive for large models and datasets, where exhaustive FI is infeasible and existing statistical approaches quickly become too time-consuming to be practical.

\subsubsection{Scalability (RepVGG and DeiT on CIFAR-10/ImageNet)}
\begin{figure}[!t]
  \centering
  \includegraphics[width=.75\columnwidth]{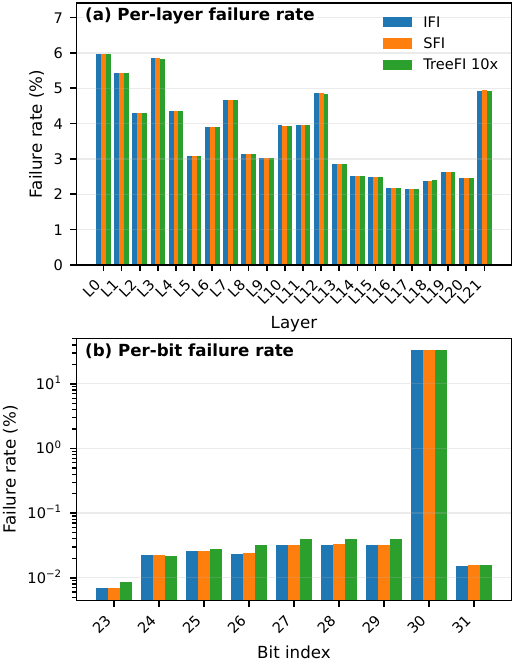}
  \vspace{-5pt}
  \caption{RepVGG failure rates: (a) per-layer over bits 23--31 and (b) per-bit over layers, for IFI, SFI, and TreeFI-$10\times$ (b uses log scale).}
  \label{fig:repvgg_marginals}
  \vspace{-10pt}
\end{figure}
We now evaluate TreeFI beyond the ResNet8 benchmark, where an exhaustive FI reference is no longer practical. In this setting, we compare TreeFI with SFI and IFI using two complementary criteria: the number of required injections and the resulting per-layer and per-bit failure-rate trends.

For RepVGG-A0, \autoref{fig:repvgg_marginals} compares the mean failure rates per layer and per bit obtained with IFI, SFI, and TreeFI-$10\times$. TreeFI-$10\times$ follows the same main per-layer and per-bit failure-rate trends as the other methods, despite using far fewer injections, as shown in \autoref{tab:inj_counts_act}. The runtime results are also consistent with the theoretical reduction in the number of required injections. On H100, the full activation-fault campaign on RepVGG-A0 requires 33\,h\,47\,min with IFI, 26\,h\,46\,min with SFI, and only 1\,h\,41\,min with TreeFI-$10\times$. These measured runtime reductions correspond to about $20.1\times$ versus IFI and $15.9\times$ versus SFI. This closely matches the theoretical reductions reported in \autoref{tab:inj_counts_act}, where TreeFI-$10\times$ requires about $21.0\times$ fewer injections than IFI and $17.5\times$ fewer injections than SFI. This agreement indicates that the wall-clock gains of TreeFI are directly explained by the reduction in the number of injections.

For DeiTs, exhaustive FI is infeasible, and IFI is also not tractable for a full-model analysis in our setup. Since the results for all DeiT versions are similar, we report only the results obtained for DeiT-Tiny.
We analyzed the first layer of DeiT-Tiny with SFI, IFI, and TreeFI, executed on the same H100 GPU. On the first DeiT-Tiny layer, IFI requires 36\,h\,43\,min, SFI requires 40\,h\,55\,min, and TreeFI-$10\times$ requires only 20\,min. These measured runtime reductions correspond to about $110\times$ versus IFI and $123\times$ versus SFI. They are again close to the theoretical reduction in the number of injections on that same layer: TreeFI-$10\times$ requires only 1{,}042 injections, compared with 124{,}089 for IFI and 139{,}778 for SFI, that is, about $119\times$ and $134\times$ fewer injections, respectively.

\begin{figure}[!t]
  \centering
  \includegraphics[width=.75\columnwidth]{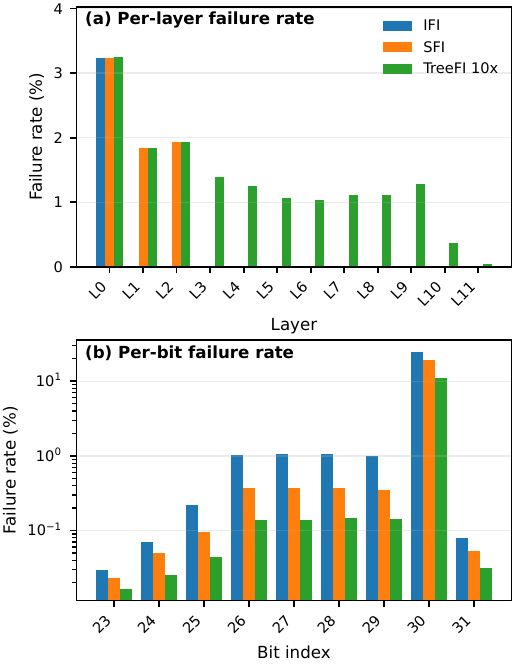}
  \vspace{-5pt}
    \caption{DeiT-Tiny failure rates: (a) per-layer over bits 23--31 and (b) per-bit over layers, for SFI and TreeFI-$10\times$ (b uses log scale).}
  \label{fig:deit_tiny_marginals}
  \vspace{-10pt}
\end{figure}

Moreover, we performed a full DeiT-Tiny activation reliability analysis using TreeFI-$10\times$, which took 6\,h\,42\,min. By contrast, SFI required 111\,h\,30\,min to complete only the first three layers of DeiT-Tiny. The failure-rate results reported in \autoref{fig:deit_tiny_marginals} show that, on the layers where SFI could be completed, TreeFI-$10\times$ follows the same main per-layer and per-bit trends while requiring a much smaller number of injections. As discussed above for the lower MAE obtained with TreeFI (Figure~\ref{fig:resnet8_mae}(a)), the differences in the reported per-layer failure rates in \autoref{fig:deit_tiny_marginals} mainly come from the reweighting step used by TreeFI, which combines interval-level results according to the empirical value distribution. To confirm this explanation, we applied the same reweighting to the SFI results, which led to failure-rate estimates that matched those of TreeFI-$10\times$. This highlights the practical advantage of TreeFI for larger transformer models, where baseline statistical FI becomes extremely costly, enabling statistically accurate and faster estimation.

\subsubsection{Hyperparameter ablation study (ResNet8 activations)}
\label{sec:ablation}
To better justify the choice of the sigmoid allocation parameters, we performed an ablation study on the ResNet8 activation benchmark, where exhaustive FI provides a ground-truth reference.
We evaluated $88$ combinations formed by sharpness values
$s \in \{1,\allowbreak 2,\allowbreak 3,\allowbreak 4,\allowbreak 6,\allowbreak 8,\allowbreak 16,\allowbreak 32\}$
and bias multipliers
$\gamma \in \{1,\allowbreak 2,\allowbreak 5,\allowbreak 10,\allowbreak 20,\allowbreak 50,\allowbreak 200,\allowbreak 2000,\allowbreak 10000,\allowbreak 20000,\allowbreak 50000\}$.
For each configuration, we measured the difference in the failure rates of TreeFI and exhaustive FI and retained the configurations in which the difference remained within the target $1\%$ error margin at the layer--bit level.
Out of the 88 tested combinations, {60} remained within the target error margin, indicating that TreeFI is not overly sensitive to moderate changes in the examined hyperparameters.
Overall, we observe a clear trend with respect to sharpness.
For all configurations with $s \geq 16$, only three biases satisfied the $1\%$ criterion, indicating that excessively sharp sigmoid mappings over-concentrate the fault injection allocation and degrade failure rate quality.
By contrast, for all configurations with $s \leq 3$, every tested bias remained within the target $1\%$ error margin, showing that the proposed methodology is robust to the choice of the bias, as long as the sharpness remains moderate.
\begin{figure}[!t]
  \centering
  \includegraphics[width=.85\columnwidth]{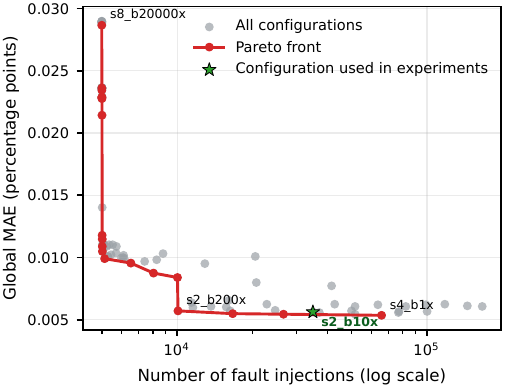}
  \vspace{-5pt}
    \caption{ResNet8 activation hyperparameter ablation for TreeFI. Each point represents a $(s,\gamma)$ configuration (x-axis: total injections; y-axis: MAE vs. exhaustive FI). Configurations meeting the $1\%$ layer$\times$bit error criterion define the feasible region. The main activation setting ($s=2$, $\gamma=10$) is highlighted.}
  \label{fig:ablation_pareto}
  \vspace{-5pt}
\end{figure}
\autoref{fig:ablation_pareto} depicts the trade-off between TreeFI's MAE relative to exhaustive fault injection (y-axis) and the total number of injections (x-axis).
The figure shows that the TreeFI configuration used in our experiments (TreeFI-$10\times$ with $s=2$) lies in a conservative region of the trade-off: it remains within the error margin, while it achieves a substantial reduction in injections.
Overall, this ablation study indicates that TreeFI is not highly sensitive to the precise bias choice when $s$ is moderate, and very large sharpness values should be avoided.
The outcome of this study supports the use of $s=2$ as a robust default value for sharpness, leading to a rather conservative TreeFI configuration in the experiments.
This suggests that the sharpness parameter is more critical than the bias multiplier: moderate sharpness values remain robust across a wide range of $\gamma$, whereas overly large sharpness values make the allocation too selective. Thus, we recommend starting from moderate sharpness and only increasing selectivity when preliminary experiments justify it.

\subsection{Faults in weights}
We additionally study bit-flip faults occurring in model weights. In contrast to activations, weight values are input-independent. Therefore, the offline characterization is computed directly from the model parameters without any dataset pass.
We completed all-layer ResNet8 campaigns, 10-layer RepVGG-A0 campaigns, and 6-layer DeiT-Tiny campaigns for IFI, SFI, and TreeFI-$20\times$.

On ResNet8 weights, TreeFI-$20\times$ required 79{,}426 injections and 1\,h\,30\,min, compared with 753{,}050 injections and 14\,h\,34\,min for SFI, and 1{,}043{,}629 injections and 20\,h\,06\,min for IFI. This corresponds to about $9.5\times$ fewer injections than SFI and $13.1\times$ fewer injections than IFI. On RepVGG-A0, TreeFI-$20\times$ required 62{,}063 injections and 1\,h\,55\,min, compared with 1{,}108{,}540 injections and 34\,h\,27\,min for SFI, and 1{,}560{,}584 injections and 49\,h\,03\,min for IFI. This corresponds to about $17.9\times$ and $25.1\times$ fewer injections, respectively. On DeiT-Tiny, TreeFI-$20\times$ required 48{,}465 injections and 15\,h\,46\,min, compared with 310{,}446 injections and 100\,h\,55\,min for SFI, and 694{,}557 injections and 227\,h\,27\,min for IFI. This corresponds to about $6.4\times$ fewer injections than SFI and $14.3\times$ fewer injections than IFI.
The corresponding wall-clock speedups are $9.7\times$, $18.0\times$, and $6.4\times$ over SFI, and $13.4\times$, $25.6\times$, and $14.4\times$ over IFI, for ResNet8, RepVGG-A0, and DeiT-Tiny, respectively. Averaged over these three executed scopes, TreeFI-$20\times$ is $11.4\times$ faster than SFI and $17.8\times$ faster than IFI. These values closely follow the average injection-count reductions of $11.2\times$ and $17.5\times$, confirming that the reduction in executed faults translates directly into campaign-time savings.

The resulting failure-rate estimates also remain close across methods. \autoref{tab:weights_mae} reports both per-layer MAE, computed after aggregating over bits, and all-bits MAE, computed after aggregating over the evaluated layers. All MAEs between TreeFI-$20\times$ and the two baselines remain far below the target 1\% error margin. The largest difference is 0.0437\% for the RepVGG-A0 all-bits comparison with IFI, more than $20\times$ smaller than the target margin. The DeiT-Tiny comparisons remain similarly small despite its substantially longer per-injection execution time.

\begin{table}[!tb]
\centering
\caption{MAE between TreeFI-$20\times$ and the weight FI baselines.}
\vspace{-5pt}
\label{tab:weights_mae}
\setlength{\tabcolsep}{4pt}
\scalebox{0.865}{%
\begin{tabular}{llrr}
\toprule
Model & Evaluation & vs. SFI~\cite{ruospoAssessingConvolutionalNeural2023a} & vs. IFI~\cite{ruospoEffectiveIterativeStatistical2025a} \\
\midrule
ResNet8 & Per-layer & 0.0000\% & 0.0367\% \\
        & All bits  & 0.0062\% & 0.0363\% \\
\midrule
RepVGG-A0 & Per-layer & 0.0190\% & 0.0340\% \\
          & All bits  & 0.0259\% & 0.0437\% \\
\midrule
DeiT-Tiny & Per-layer & 0.0183\% & 0.0250\% \\
          & All bits  & 0.0243\% & 0.0306\% \\
\bottomrule
\end{tabular}
}
\vspace{-5pt}
\end{table}

\autoref{tab:inj_counts_w} summarizes the executed weight campaigns: all layers for ResNet8, 10 layers for RepVGG-A0, and 6 layers for DeiT-Tiny. TreeFI-$20\times$ requires fewer injections than both SFI and IFI in every evaluated model. The reductions are smaller than for activation faults because weight values vary less, leaving less structure for TreeFI to exploit. Even so, TreeFI-$20\times$ reduces the number of weight injections by $6.4\times$--$17.9\times$ compared with SFI and by $13.1\times$--$25.1\times$ compared with IFI, demonstrating substantial gains across all three models. Taken together, TreeFI-$20\times$ reduces the weight campaign by at least $6.4\times$, while its largest MAE relative to either baseline is only 0.0437\%, showing that the efficiency gains do not produce materially different failure-rate estimates.

\begin{table}[!tb]
\centering
\caption{Number of injections in the executed \textbf{weight} FI campaigns.}
\vspace{-5pt}
\label{tab:inj_counts_w}
\setlength{\tabcolsep}{2pt}
\resizebox{0.67\columnwidth}{!}{
\begin{tabular}{lrrr}
\hline
Method & \textbf{ResNet8} & \textbf{RepVGG-A0} & \textbf{DeiT-Tiny} \\
Scope & All layers & 10 layers & 6 layers \\
\hline
SFI \cite{ruospoAssessingConvolutionalNeural2023a} & 753{,}050 & 1{,}108{,}540 & 310{,}446 \\
IFI \cite{ruospoEffectiveIterativeStatistical2025a} & 1{,}043{,}629 & 1{,}560{,}584 & 694{,}557 \\
\hline
TreeFI 20$\times$ & 79{,}426 & 62{,}063 & 48{,}465 \\
\hline
\end{tabular}
}
\vspace{-5pt}
\end{table}

\section{Conclusions and discussion}

We presented TreeFI, a value-aware fault-injection methodology for efficient reliability evaluation in neural networks. TreeFI partitions the value distribution into intervals, derives an interval risk score from the severity of bit flips within each interval, and uses this information to assign fault injections across intervals. The final reliability estimate is then computed from the measured fault-injection outcomes. In this way, TreeFI targets a desired confidence level and error margin while avoiding a bit-level, value-uniform sampling strategy.
On the evaluated benchmarks, TreeFI substantially reduces the number of required injections compared with state-of-the-art statistical FI approaches, while keeping the estimation error within the target margin when validated against an exhaustive FI reference. The results further show that TreeFI preserves the main per-layer and per-bit failure-rate trends in larger activation-fault experiments, making it useful even when exhaustive validation is impractical. For weight faults, we validate the approach experimentally on CNN models and a 6-layer DeiT-Tiny subset.
Our experimental study focuses on FP32 bit flips, which remain common in reliability studies and provide a clear setting for evaluating the proposed methodology. However, the method itself is more general: it only requires a fault model that defines how a bit flip modifies a value, together with a value distribution that can be collected and partitioned into intervals. We therefore expect that the same interval-learning and stratified-allocation principle can be extended to other numerical formats, such as FP16, bfloat16, or quantized integer representations. Validating this experimentally is an important direction for future work.
Overall, TreeFI provides an effective compromise between estimation cost and estimate quality, and is particularly attractive for large-scale fault-injection studies where exhaustive evaluation is infeasible.

\section{Reproducibility}

The TreeFI implementation and experimental artifacts are available at \url{https://gitlab.inria.fr/nbires/treefi}. The repository's \texttt{README.md} provides installation instructions and ready-to-run activation and weight examples, while \texttt{docs/reproducibility.md} explains how to reproduce the paper's results.

\section*{Acknowledgment}
This work was supported by the French government under the France 2030 investment plan, as part of the PEPR Intelligence Artificielle (AdaptING project, grant number ANR-23-PEIA-0009). 
The experiments were performed using SLICES-FR HPC, supported by Inria and partner institutions.
% Test Test
%MT commented for blind review

\clearpage
\printbibliography

\end{document}